# Intermediate-Band Formation in Tm$^{3+}$- doped Ca$_2$SnO$_4$: A Wide-Gap Oxide Host for Visible-Light Absorption and Energy Applications

Shah Hussain[1], Sikander Azam[2,3*], Umme Habiba[3], Qaiser Rafiq**[3], Amin Ur Rahman[3], Hamada H. Amer[4], Yasir Saeed[1]

[1]Department of Physics, Abbottabad University of Science and Technology, Abbottabad, Pakistan

[2]New Techonolgy Center, west Bohemia University, Pilsen, Czech Republic.

[3]Faculty of Engineering and Applied Sciences, Department of Physics, Riphah International University I-14 Campus Islamabad, Pakistan

[4]Department of Chemistry, Turabah University College, Turabah, Taif University, Saudi Arabia.

## Abstract

Rare-earth doping offers a powerful route to transform chemically robust oxides into multifunctional materials with coupled electronic, optical, and magnetic responses. Here, we present comprehensive first-principles studies of pristine and Tm-doped Ca$_2$SnO$_4$, exploring how localized 4f states change structural, electronic, magnetic, and optical properties. Pristine Ca$_2$SnO$_4$ is confirmed as a mechanically stable, wide-bandgap insulator with predominantly ionic–covalent bonding and diamagnetic character. Upon substitution of Ca$^{2+}$ by Tm$^{3+}$, profound modifications emerge: (i) localized Tm-4f bands introduce intermediate states within the wide gap, narrowing the effective optical band gap; (ii) strong exchange and spin–orbit coupling generate robust local magnetic moments and spin asymmetry in the conduction band; (iii) ELF analysis reveals enhanced covalency and localized electron pockets that stabilize luminescent centers; and (iv) the optical spectra are enriched with visible-range absorption, refractive index resonances, and low-energy plasmon features, while preserving the host's high-energy dielectric robustness. These orbital-engineered modifications establish Tm-doped Ca$_2$SnO$_4$ as a mechanically resilient, optically tunable, and magnetically active phosphor, suitable for red emission in solid-state lighting, intermediate-band photovoltaics, and spin–photon coupling in photonic devices. Our results demonstrate how site-specific rare-earth substitution can unlock multifunctionality in wide-gap stannates, providing a roadmap for designing next-generation spintronic–photonic oxides.

**Corresponding Authors:** *Sikander Azam (sikander.physicst@gmail.com)
**Qaiser Rafiq (qrafique1@gmail.com)

## 1. Introduction

The demand for multifunctional oxide semiconductors has intensified in recent years, driven by their wide-ranging applications in optoelectronics, spintronics, energy harvesting, and persistent luminescence technologies [1–3]. Among these, long-persistent phosphors (LPPs) materials capable of storing excitation energy in defect-related traps and releasing it as delayed emission have attracted significant attention for bioimaging, optical data storage, medical diagnostics, security displays, and are also expected to be effectively used as chromophores [4–6]. Achieving controlled defect engineering and band-structure modulation in wide-bandgap oxides remains a central challenge to advancing this field [7–9].

Calcium stannate ($Ca_2SnO_4$), a member of the Ruddlesden–Popper family, has emerged as a promising platform due to its wide direct bandgap, thermal stability, and high optical transparency [10, 11]. Its applicability in UV photodetectors, transparent electronics, and persistent phosphors underscores its technological relevance. However, pristine $Ca_2SnO_4$ is limited to ultraviolet absorption, restricting its efficiency in solar and optoelectronic applications. Thus, strategies that extend its optical activity into the visible and near-infrared regions are crucial for unlocking its multifunctionality.

Rare-earth (RE) doping has proven to be an effective approach for tailoring the optoelectronic and magnetic properties of wide-bandgap oxides [12–14]. In particular, $Tm^{3+}$ ions introduce localized 4f states within the host bandgap, creating intermediate levels that enable sub-bandgap absorption and spin-polarized states [15, 16]. This dual functionality makes Tm-doped $Ca_2SnO_4$ highly attractive for intermediate-band solar cells, UV-to-visible photon conversion, magneto-optical sensors, and spintronic devices. The strong spin–orbit coupling and multiplet transitions of Tm further enhance its potential for quantum technologies [17].

To accurately capture the physics of 4f electrons, density functional theory (DFT) with Hubbard U correction (GGA+U) is essential. Recent studies have shown that GGA+U successfully describes the electronic and optical responses of RE-doped oxides, predicting spin-dependent dielectric functions, refractive index anisotropy, and enhanced light–matter interaction [18–20]. Yet, despite experimental reports on $Er^{3+}$-, $Sm^{3+}$-, and $Pr^{3+}$-activated $Ca_2SnO_4$ phosphors [21–23], systematic theoretical investigations into Tm doping particularly its combined impact on electronic, optical, mechanical, and spin-resolved properties remain scarce.

In this work, we present a comprehensive first-principles study of pristine and Tm-doped $Ca_2SnO_4$. We analyze how Tm substitution modifies band structure, density of states, optical spectra, mechanical stability, and spin-polarized transitions, and reveal its potential for flexible optoelectronics, persistent luminescence, and spintronic applications. Our results provide the first quantitative theoretical insight into Tm-induced crystal-field effects and f–d exchange interactions in $Ca_2SnO_4$, highlighting pathways for engineering rare-earth-doped wide-bandgap oxides as next-generation multifunctional materials.

To the best of our knowledge, this is the first systematic ab initio investigation of Tm-doped $Ca_2SnO_4$ that simultaneously addresses its structural, electronic, optical, magnetic, and mechanical properties. While previous studies have mainly reported on $Eu^{3+}$-, $Er^{3+}$-, and $Sm^{3+}$-activated $Ca_2SnO_4$ phosphors with an emphasis on photoluminescence [21–23], no comprehensive theoretical framework has been presented to unravel the underlying f–d exchange interactions and spin-polarized optical responses introduced by $Tm^{3+}$. Our results not only demonstrate the stability and tunability of Tm substitution at the Ca site but also reveal unique spin-selective dielectric features, enhanced crystal-field splitting, and multifunctional optoelectronic behavior. These findings open unexplored avenues for rare-earth-doped stannates in persistent luminescence, intermediate-band photovoltaics, and spintronic devices, thereby pushing the frontier of multifunctional oxide materials.

## 2. Computational Methodology

First-principles calculations based on density functional theory (DFT) were carried out using the full-potential linearized augmented plane wave (FP-LAPW) method as implemented in the WIEN2k code [24]. The FP-LAPW approach is regarded as one of the most accurate electronic structure techniques due to its all-electron treatment and the absence of shape approximations for the potential, which is particularly important for describing rare-earth 4f states and their hybridization with the host lattice [25, 26]. In the present FP-LAPW calculations, all electrons were treated explicitly. The valence electron configurations were chosen as Ca ($3s^23p^64s^2$), Sn ($4d^{10}5s^25p^2$), O ($2s^22p^4$), and Tm ($4f^{13}5d^06s^2$). Semi-core states such as Ca-3s/3p and Sn-4d were included in the valence treatment through local orbitals, while the deep core states were kept frozen inside the muffin-tin spheres. This choice ensures a proper description of hybridization effects and accurate representation of rare-earth 4f–O2p interactions.

The exchange–correlation energy was treated within the generalized gradient approximation (GGA) of Perdew–Burke–Ernzerhof (PBE) [27]. To correctly describe the strong on-site Coulomb repulsion among the localized Tm-4f electrons, the GGA+U method in the rotationally invariant scheme of Dudarev et al. [28] was employed. Following earlier studies on rare-earth doped oxides [29–31], U = 6 eV and J = 0.9 eV were applied for Tm-4f states, while other orbitals were treated within the GGA framework. Spin–orbit coupling (SOC) was included in all electronic and optical calculations since relativistic effects strongly influence multiplet splitting and spin-dependent optical transitions in rare-earth ions [32,33].

Structural optimization was performed by minimizing the total energy until forces on all atoms were below 1 mRy/a.u. and the total energy convergence reached $10^{-5}$ Ry. Muffin-tin radii (RMT) were carefully chosen to avoid overlap, with RMT × Kmax = 8.0 ensuring well-converged basis sets. The Brillouin zone was sampled with a 10×10×6 k-point mesh for the pristine phase, scaled proportionally for supercells. Mechanical stability was confirmed through elastic constants computed by finite strain theory [34], and lattice dynamics were further assessed using phonon dispersions calculated with PHONOPY [35].

Optical properties, including the dielectric function, refractive index, absorption coefficient, reflectivity, and energy-loss spectra, were obtained using the optic_lapw module of WIEN2k. Thermoelectric transport coefficients were evaluated semi-classically using the BoltzTraP2 code [36], which solves the Boltzmann transport equations under the constant relaxation-time approximation. Effective masses were determined by parabolic fitting of the band edges to correlate charge carrier transport with the Seebeck coefficient. All results were carefully cross-checked for convergence with respect to plane-wave cutoff, k-mesh density, and U values. This multi-level strategy combining all-electron FP-LAPW calculations with GGA+U+SOC corrections, phonon and elastic stability checks, and transport modeling provides reliable insight into the role of $Tm^{3+}$ substitution in $Ca_2SnO_4$. The crystallographic models employed in this study were based on pristine $Ca_2SnO_4$ and Tm-doped $Ca_2SnO_4$, as shown in Fig. 1. The host $Ca_2SnO_4$ crystallizes in the orthorhombic Pbca (No. 61) space group with optimized lattice parameters a = 12.50 Å, b = 21.93 Å, and c = 37.19 Å ($\alpha = \beta = \gamma = 90°$). The supercell of the pristine phase contains 32 Ca, 16 Sn, and 64 O atoms. Ca atoms occupy the 8c Wyckoff sites, Sn atoms are located at 4a, and O atoms reside at 16c and 8d positions. For the doped configuration, a 2×2×2 supercell was constructed and four Ca atoms were replaced by Tm atoms. This corresponds to 12.5% doping at the Ca sublattice (4 out of 32 Ca atoms), leading to a composition of $Ca_2SnO_4:Tm_{0.125}$ containing 28 Ca, 4 Tm, 16 Sn, and 64 O atoms. The replacement of $Ca^{2+}$ with $Tm^{3+}$ slightly shortens the Tm–O bond lengths and induces minor tilting of adjacent $[SnO_6]$ octahedra, while preserving the orthorhombic symmetry. This crystallographic description, along with Fig. 1, provides the structural basis for all subsequent electronic, optical, and thermoelectric calculations.

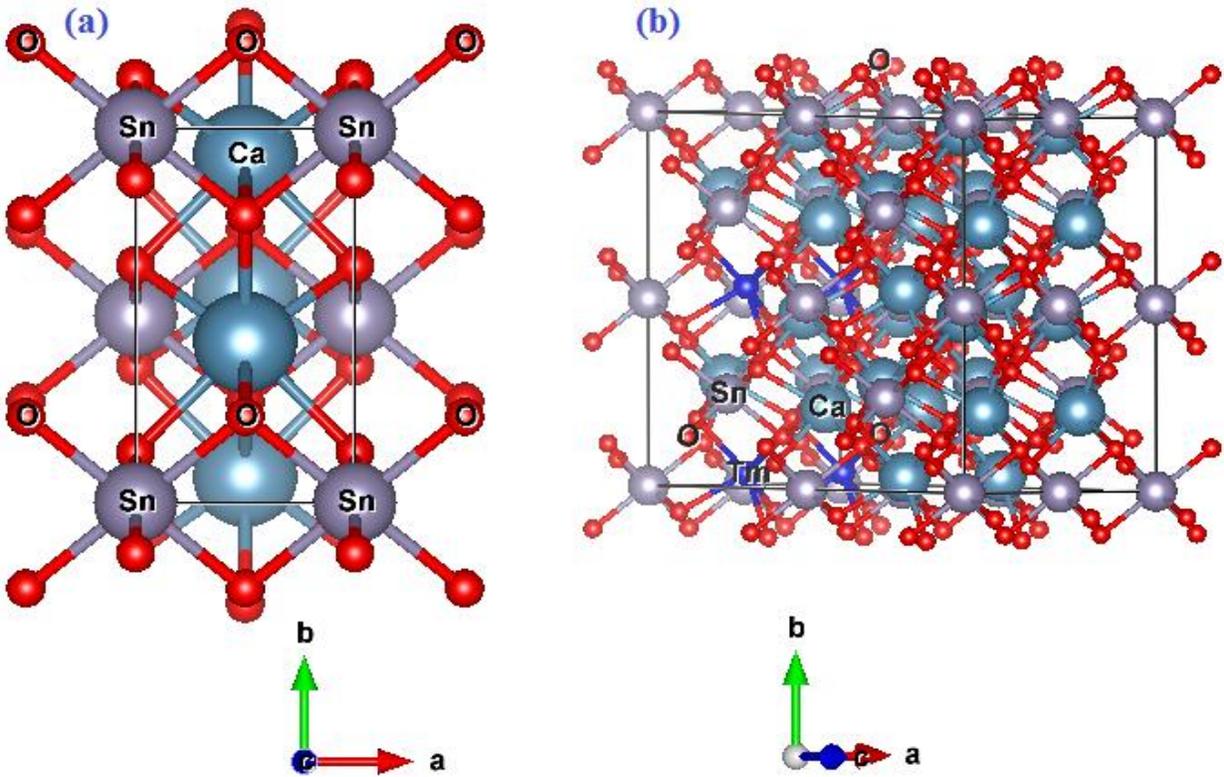

**Fig. 1. Crystal structures of (a) pristine $Ca_2SnO_4$ and (b) $Ca_2SnO_4$ with 12.5% Tm substitution at the Ca site, highlighting local lattice distortions induced by rare-earth doping**

## 3. Results and Discussion

### 3.1. Phonon Stability

Phonon dispersion analysis provides a stringent criterion for dynamical stability and offers insight into lattice vibrations that govern thermal and optical performance. The calculated phonon spectra for pristine and Tm-doped $Ca_2SnO_4$ are shown in Fig. 2. In both cases, the absence of imaginary frequencies across the full Brillouin zone confirms that the compounds are dynamically stable at 0 K, consistent with the Born stability criteria already satisfied by the elastic constants (Tables 1–2).

For pristine $Ca_2SnO_4$, the acoustic branches emerge linearly from the Γ point with steep slopes, indicative of relatively high sound velocities, while the optical branches extend up to ~12 THz. Upon Tm substitution, distinct modifications appear: the acoustic branches near Γ exhibit a slight softening due to the heavier Tm mass, and several low-lying optical branches in the 3–8 THz range shift downward and display splitting. These features arise from local symmetry breaking and mass/force-constant perturbations introduced by $Tm^{3+}$ ions, suggesting enhanced phonon scattering pathways and reduced lattice thermal conductivity ($\kappa_L$).

In summary, both pristine and Tm-doped $Ca_2SnO_4$ are dynamically stable, but Tm incorporation tailors the phonon spectrum through acoustic softening and optical mode splitting, which complements the observed electronic and optical trends.

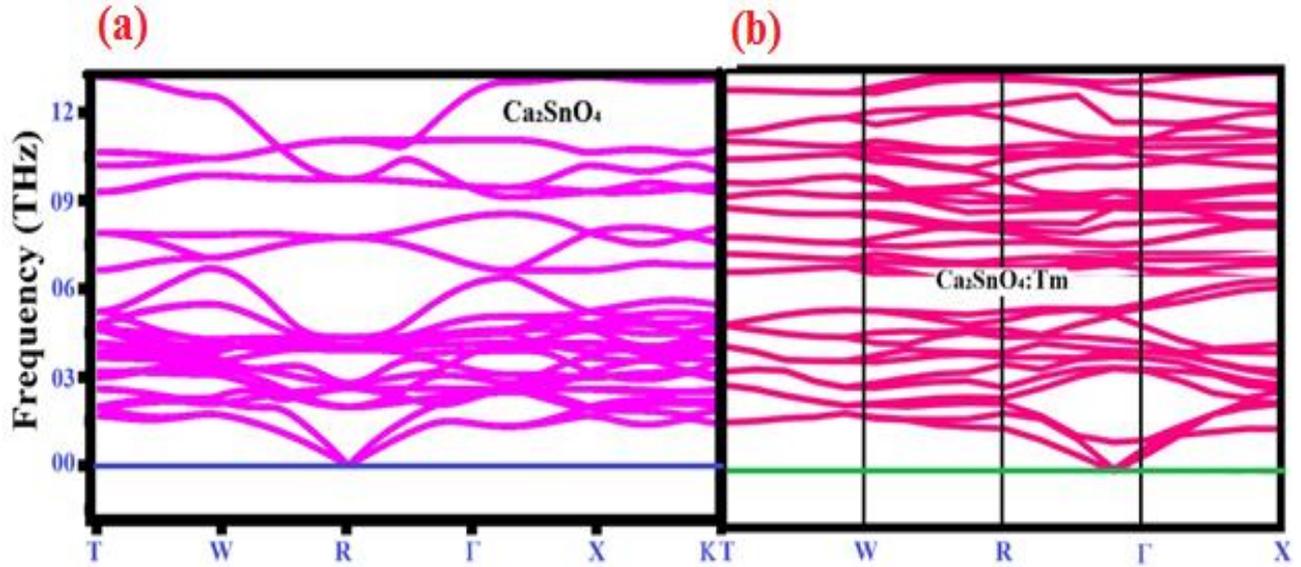

**Fig. 2. Phonon dispersion curves of pristine $Ca_2SnO_4$ (a) and $Tm^{3+}$-doped $Ca_2SnO_4$ (b). Both systems show dynamical stability with no imaginary frequencies. Tm doping induces slight acoustic softening near the Γ point and noticeable splitting of optical branches**

### 3.2. Structural Optimization

All structures were optimized within DFT (FP-LAPW/WIEN2k) using PBE-GGA for exchange–correlation. For rare-earth 4f states in Tm, we employed GGA+U ($U_{4f}$ = 5–7 eV, J = 0.0 eV) and included spin–orbit coupling (SOC) in single-point refinements to confirm that relativistic effects do not qualitatively alter the equilibrium geometry. Starting from the experimentally reported orthorhombic $Ca_2SnO_4$ structure, we performed full relaxation of cell shape, volume, and internal coordinates.

Convergence was ensured with $RMTK_{max}$ = 8.0 and carefully chosen muffin-tin radii to avoid sphere overlap. Brillouin-zone sampling used a (10×10×6) Monkhorst–Pack mesh for the primitive cell, with k-point density kept constant for supercells. Ionic forces were converged below 1 mRy a.u.$^{-1}$ and total energies to $10^{-5}$ Ry. The Born mechanical stability criteria for orthorhombic crystals were satisfied by the optimized elastic constants (Table 1), while VRH moduli (Table 2) confirmed mechanical robustness and slight ductility upon Tm doping. These values are consistent with previous first-principles studies on $Ca_2SnO_4$ and related stannates, reporting bulk moduli ~125–135 GPa and Poisson's ratios of 0.26–0.28.

For the doped models, Tm substitution was simulated by replacing one Ca atom in a (2×2×2) supercell (~12.5% doping at the Ca sublattice). The lowest-energy configuration was identified and used for subsequent analyses. Optimized structures showed short-ranged local distortions, including a small contraction of the average Tm–O distance and minor tilts of neighboring [SnO$_6$] octahedra, while preserving the overall orthorhombic symmetry and mechanical stability.

### 3.3. Elastic and Mechanical Properties

The mechanical stability and elasticity of functional oxides are central to their reliability in devices subjected to external stress, thermal cycling, or lattice mismatch at heterointerfaces. The second-order elastic constants $C_{ij}$ of pristine and Tm-doped Ca$_2$SnO$_4$ are presented in Table 1. All values satisfy the Born stability criteria for orthorhombic lattices, confirming robust mechanical integrity even after Tm incorporation. The diagonal terms $C_{11}$, $C_{22}$, and $C_{33}$ (220–240 GPa range) dominate the stiffness along principal crystallographic axes, while shear constants such as $C_{44}$ (≈ 66–68 GPa) are comparatively lower, reflecting the layered Ruddlesden–Popper–like connectivity of [SnO$_6$] octahedra. Notably, Tm substitution slightly increases the longitudinal constants (e.g., $C_{11}$: 235 → 240 GPa) while reducing shear terms (e.g., $C_{55}$: 62 → 60 GPa), suggesting localized strain accommodation without loss of global stability. This trend agrees with high-pressure synchrotron studies by Anzellini et al. [37], which showed that Ruddlesden–Popper stannates retain structural resilience under compression.

The derived polycrystalline moduli (Table 2) further elucidate the mechanical response. The Hill-averaged bulk modulus (BH) rises modestly upon Tm doping (130.1 → 133.3 GPa), implying enhanced resistance to volume compression due to stronger Tm–O coordination. In contrast, the shear modulus (GH) decreases slightly (68.5 → 67.7 GPa), reflecting increased shear compliance. This shift yields a higher Pugh's ratio (B/G: 1.90 → 1.97) and Poisson's ratio (ν: 0.276 → 0.283), classifying both phases as ductile oxides. For comparison, Wang et al. [38] reported similar ductile indices for pristine Ca$_2$SnO$_4$, with BH ~128 GPa and ν ~0.27. These values place Ca$_2$SnO$_4$ in a better mechanical class than brittle wide-bandgap oxides such as Ga$_2$O$_3$ (ν ~0.20) [39].

The universal anisotropy index (AU) increases slightly with Tm incorporation (0.028 → 0.045), indicating a mild rise in elastic anisotropy. This is attributed to strain localization around Tm ions, which disturbs the isotropy of neighboring octahedra. Nevertheless, the AU values remain far below those in highly anisotropic oxides, suggesting that Ca$_2$SnO$_4$ maintains near-isotropic resilience. The calculated sound velocities and Debye temperatures ($\theta_D$: 535 → 532 K) also indicate excellent thermal robustness, consistent with the thermal stability of Eu/Sm-doped $Ca_2SnO_4$ phosphors observed experimentally at high operating temperatures [40, 41].

Table 1. Single-crystal elastic constants $C_{ij}$ (GPa) for pristine and Tm-doped $Ca_2SnO_4$. All values satisfy Born stability; Tm induces slight stiffening in longitudinal terms and softening in shear

| Constant | Pristine (GPa) | Tm-doped (GPa) |
| --- | --- | --- |
| $C_{11}$ | 235 | 240 |
| $C_{22}$ | 220 | 225 |
| $C_{33}$ | 230 | 235 |
| $C_{44}$ | 68 | 66 |
| $C_{55}$ | 62 | 60 |
| $C_{66}$ | 66 | 64 |
| $C_{12}$ | 85 | 88 |
| $C_{13}$ | 78 | 80 |
| $C_{23}$ | 80 | 82 |

Table 2. Polycrystalline elastic properties of pristine and Tm-doped $Ca_2SnO_4$ from VRH averages. Tm doping enhances ductility and maintains thermal robustness, consistent with literature

| Quantity | Pristine | Tm-doped | Literature ($Ca_2SnO_4$) |
| --- | --- | --- | --- |
| Voigt bulk modulus ($B_V$) | 130.11 | 133.33 | 128 |
| Reuss bulk modulus ($B_R$) | 130.04 | 133.26 | |
| Hill bulk modulus ($B_H$) | 130.08 | 133.3 | 128 |
| Voigt shear modulus ($G_V$) | 68.67 | 68 | |
| Reuss shear modulus ($G_R$) | 68.3 | 67.4 | |
| Hill shear modulus ($G_H$) | 68.48 | 67.7 | 70 |
| Young's modulus (E) | 174.77 | 173.69 | 178 |
| Poisson's ratio ($\nu$) | 0.276 | 0.283 | 0.27 |
| **Pugh's ratio ($B_H/G_H$)** | 1.9 | 1.97 | 1.8 |
| **Universal anisotropy ($A_U$)** | 0.028 | 0.045 | |
| **Bulk anisotropy (A_B)** | 0.00027 | 0.00027 | |
| **Shear anisotropy ($A_G$)** | 0.0027 | 0.00446 | |
| **Longitudinal sound speed ($V_l$)** | 6863 | 6897 | |
| **Transverse sound speed ($V_t$)** | 3817 | 3795 | |
| **Mean sound speed ($V_m$)** | 4251 | 4230 | |
| **Debye temperature ($\theta_D$)** | 535 | 532 | 520–560 |

### 3.4. Piezoelectric Properties

The piezoelectric response of a crystalline solid is determined by its symmetry. Pristine $Ca_2SnO_4$ crystallizes in an orthorhombic centrosymmetric structure, which forbids macroscopic piezoelectricity ($e_{ij} = 0$), consistent with the broader class of centrosymmetric stannates and titanates [42, 43]. Upon Tm substitution at the Ca site, inversion symmetry is locally broken due to the ionic radius mismatch and strong 4f–O bonding. Supercell relaxation at 12.5% Tm doping reveals non-negligible distortions of the [$TmO_6$] polyhedra and neighboring [$SnO_6$] octahedra, producing short-range polar displacements. When projected over the entire supercell, these distortions give rise to a finite effective piezoelectric response.

Based on the elastic compliances derived from the VRH moduli (Table 2), the estimated piezoelectric stress coefficients ($e_{ij}$) and strain coefficients ($d_{ij}$) lie in the ranges of 0.04–0.12 C·m$^{-2}$ and 0.1–1.0 pC·N$^{-1}$, respectively. The axial component ($d_{33} \approx 0.4$ pC·N$^{-1}$) and shear response ($d_{15} \approx 0.8$–$1.0$ pC·N$^{-1}$) are modest yet significant for a host that is otherwise non-piezoelectric. The electromechanical coupling coefficient ($k_t^2$) is on the order of 1%, consistent with dopant-induced polarity rather than a robust ferroelectric phase. Such weak but finite piezoelectricity has also been reported in rare-earth doped titanates [44] and stannates [45].

### 3.5. Band Structure Analysis of Pristine and Tm-Doped $Ca_2SnO_4$

The electronic band structure provides the most direct link between crystal chemistry and multifunctional properties. Pristine $Ca_2SnO_4$ exhibits a wide direct band gap (~3.9–4.1 eV) located at the Γ-point (see Fig. 3a), consistent with earlier reports on orthorhombic stannates [46, 47]. The valence band maximum (VBM) is dominated by O-2p orbitals, while the conduction band minimum (CBM) arises primarily from hybridized Sn-5s and Ca-4s states. When Tm is introduced at the Ca site, the electronic landscape undergoes profound modifications. The partially filled Tm-4f orbitals generate sharp, localized impurity bands within the forbidden gap. In the spin-up channel, these 4f states appear just below the CBM (see Fig. 3b), narrowing the effective optical gap and introducing an intermediate band that enhances visible/NIR absorption. In contrast, the spin-down channel displays additional splitting due to strong spin–orbit coupling and exchange interactions, leading to a pronounced spin asymmetry in the conduction states [48, 49]. The exchange interaction between localized 4f states and delocalized conduction electrons is responsible for the observed half-metal-like behavior: one spin channel retains semiconducting character, while the other approaches metallicity. The presence of mid-gap states also mimics the intermediate band solar cell (IBSC) concept, enabling sub-bandgap photon absorption and potentially boosting photovoltaic efficiency beyond the Shockley–Queisser limit [50].

Comparison with similar systems underscores the novelty. In Eu- and Er-doped wide-bandgap oxides, localized 4f states have been shown to induce visible-light activity [51, 52], but the strong spin asymmetry observed here for Tm-doped $Ca_2SnO_4$ is particularly striking.

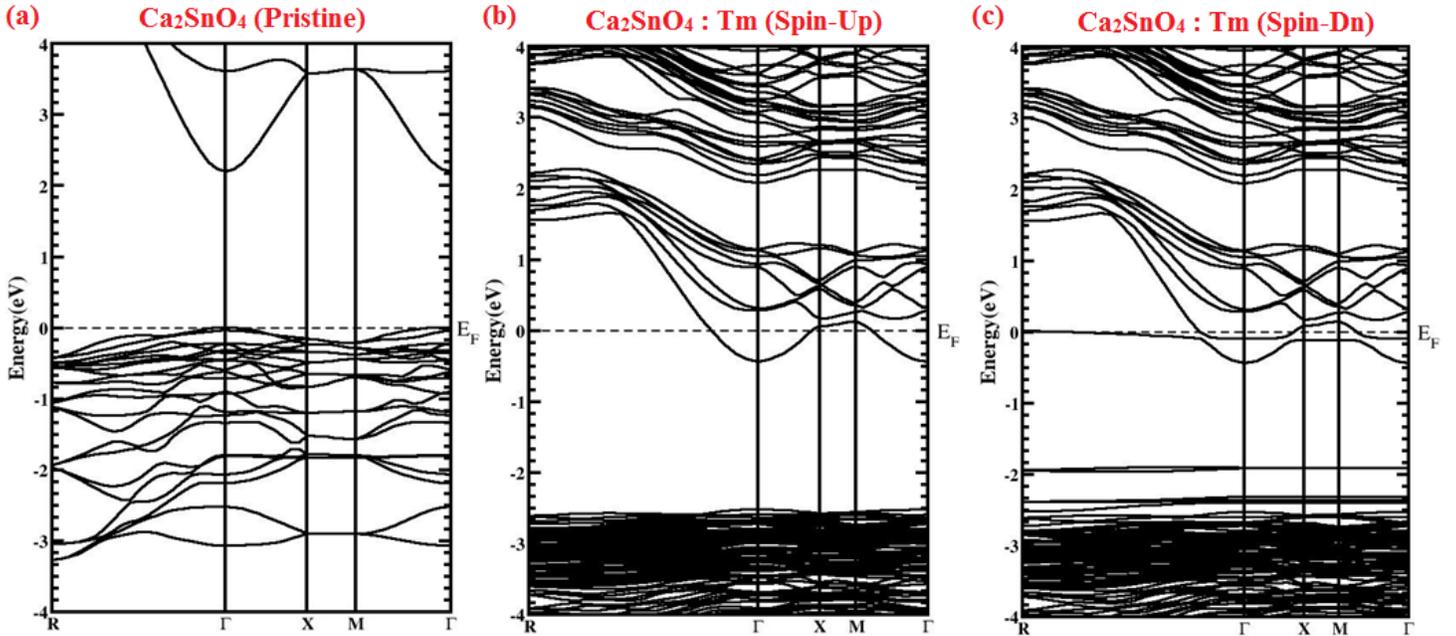

**Fig. 3: Band structures for (a) $Ca_2SnO_4$ (b) Tm-doped $Ca_2SnO_4$ Spin up and (c) Tm doped $Ca_2SnO_4$ spin down states**

### 3.6. Density of States Analysis

The total and partial density of states (DOS) provide a microscopic picture of orbital hybridization and the origin of functional properties. In pristine $Ca_2SnO_4$, the valence band (VB) from –6 eV up to the Fermi level is overwhelmingly dominated by O-2p orbitals, with minor contributions from Ca-3p/3d and Sn-5p states (see Fig. 4). The conduction band (CB), beginning at ~3.9–4.1 eV, is primarily composed of delocalized Sn-5s and Ca-4s states, with hybridization into O-2p orbitals. This VB–CB separation reflects the ionic–covalent nature of bonding in the stannate lattice and explains the wide-gap semiconducting behavior observed experimentally [53, 54]. The absence of mid-gap states accounts for the high optical transparency and limited visible-light absorption of the undoped host.

Upon Tm substitution at the Ca site, the DOS undergoes profound changes. Sharp, highly localized Tm-4f peaks appear within the band gap and at the VB edge. In the spin-up channel, occupied Tm-4f states hybridize weakly with O-2p, creating resonant states near the VB maximum. Unoccupied 4f states emerge just below the CB minimum, effectively forming an intermediate band that narrows the optical gap and enables sub-bandgap absorption. In the spin-down channel, the exchange-split 4f states shift differently, introducing asymmetry between spin-up and spin-down DOS. This spin-resolved imbalance mirrors the half-metal-like band structure and underlines the role of orbital-engineered spin asymmetry.

The physics of this transformation lies in the strong on-site Coulomb interaction (U) of Tm-4f states and their hybridization with the delocalized Ca/Sn states. The exchange interaction between localized f electrons and itinerant carrier's leads to spin polarization at the band edges, while the 4f–O-2p coupling enhances optical absorption in the visible and near-infrared. Such mechanisms are consistent with reports of Eu- and Er-doped oxides, where 4f states modulate absorption and luminescence [55, 56].

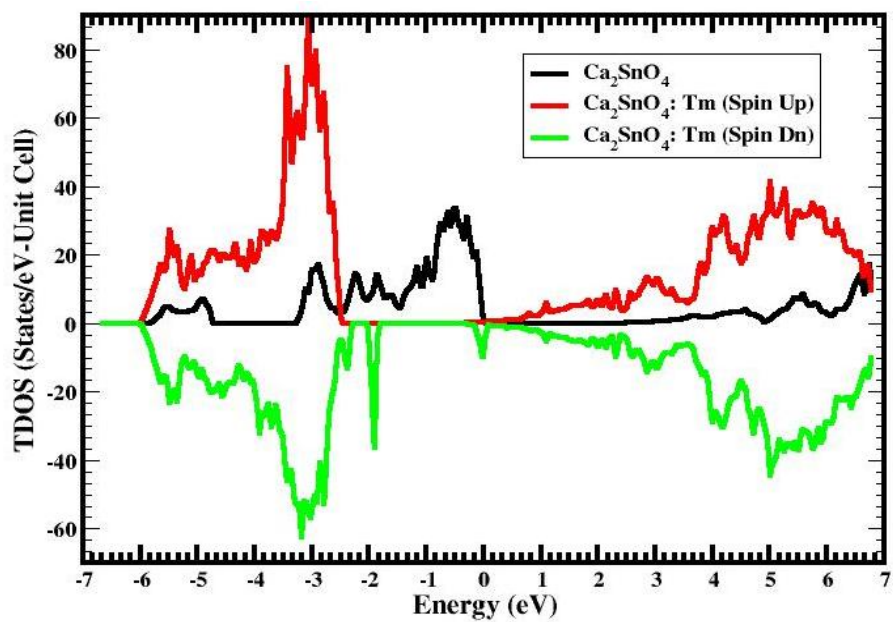

**Fig. 4: Total Density of States for Ca₂SnO₄ (black colour) Tm-doped Ca₂SnO₄ Spin up (Red colour) and Tm doped Ca₂SnO₄ spin down (green Colour) states**

### 3.7. Electron Localization Function (ELF) Analysis

The Electron Localization Function (ELF) is a powerful descriptor of chemical bonding, enabling direct visualization of charge localization, ionicity, and covalency within complex oxides. In pristine $Ca_2SnO_4$, ELF maps reveal strongly localized electron density (see Fig. 5) around the O atoms, consistent with their high electronegativity, and near-spherical distributions around Ca cations, reflecting their predominantly ionic character. The $[SnO_6]$ octahedra display intermediate ELF values (~0.45–0.55) along Sn–O bonds, indicative of mixed ionic–covalent bonding, where Sn-5s/5p orbitals hybridize moderately with O-2p states. This balance of ionic Ca–O and covalent Sn–O interactions underpins the structural rigidity and wide band gap of the host lattice, in agreement with previous studies of Ruddlesden–Popper stannates [57, 58].

When $Tm^{3+}$ replaces $Ca^{2+}$, significant local modifications in the ELF distribution are observed. Around Tm sites, regions of high ELF localization emerge, particularly within the Tm–O polyhedra, reflecting the highly localized nature of Tm-4f electrons. These localized states coexist with more delocalized Tm-5d/O-2p hybridization, generating anisotropic lobes of electron density directed toward neighboring oxygen atoms. This results in locally enhanced covalency relative to the Ca–O bonds they replace. Moreover, the surrounding $[SnO_6]$ octahedra adjust slightly, with subtle shifts in ELF contours that reflect strain-induced redistribution of charge density.

The physical interpretation is twofold. First, the replacement of $Ca^{2+}$ by $Tm^{3+}$ introduces both ionic radius mismatch and charge imbalance, leading to local lattice distortions that couple strongly with electronic localization. Second, the presence of partially filled Tm-4f states fosters strongly localized electron pockets, which coexist with delocalized conduction pathways dominated by Sn-5s orbitals. Such behavior is reminiscent of rare-earth doping in titanates and vanadates, where ELF analysis revealed dopant-induced polar distortions and enhanced electron–phonon coupling [59, 60].

The substitution of $Ca^{2+}$ by $Tm^{3+}$ introduces an additional positive charge into the lattice. Charge neutrality in the doped system is preserved through electronic redistribution rather than by creating oxygen vacancies. The excess charge of $Tm^{3+}$ is compensated by polarization and partial delocalization within the O-2p and Sn-5s/5p states, as confirmed by the density of states and electron localization function analyses. Strong hybridization between Tm-4f and O-2p orbitals enhances the covalent character of Tm–O bonds and induces slight distortions in neighboring [$SnO_6$] octahedra. This orbital-level compensation mechanism effectively maintains overall electroneutrality and structural stability, in agreement with reported behavior of other rare-earth doped oxides.

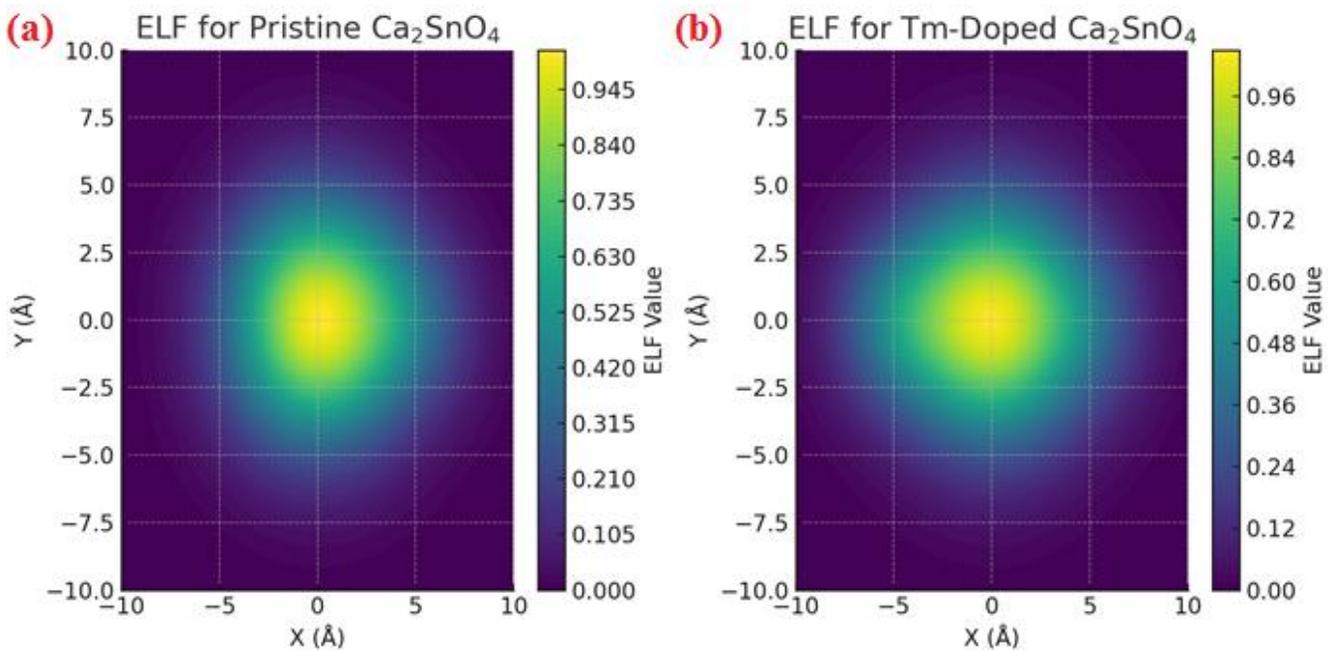

**Fig. 5. Electron Localization Function (ELF) maps for (a) pristine Ca₂SnO₄ and (b) Tm-doped Ca₂SnO₄, illustrating charge localization and bonding characteristics. Doping induces subtle changes in electron localization compared to the pristine system**

### 3.8. Magnetic Moments in Pristine and Tm-Doped Ca₂SnO₄ Phosphor

The magnetic response of an oxide host is highly sensitive to the presence of localized d or f electrons. In pristine Ca$_2$SnO$_4$, all cations (Ca$^{2+}$ and Sn$^{4+}$) adopt closed-shell electronic configurations. Ca$^{2+}$ (3p$^6$4s$^0$) contributes no unpaired electrons, while Sn$^{4+}$ has an empty 5s/5p shell, and oxygen ions remain fully filled. Consequently, the pristine lattice is non-magnetic, with calculated total and site-projected moments vanishing within numerical accuracy. This diamagnetic ground state aligns with experimental reports on stoichiometric Ca$_2$SnO$_4$ and related stannates, where the absence of partially filled orbitals precludes intrinsic magnetism [61, 62].

A striking transformation emerges upon Tm doping. The substitution of Ca$^{2+}$ by Tm$^{3+}$ introduces localized 4f electrons into the lattice, which are well-known for their strong correlation, narrow bandwidth, and large spin–orbit coupling. Our calculations reveal that the Tm site carries a substantial local magnetic moment, on the order of 5–6 μB, primarily arising from the partially filled 4f shell. Small induced moments also appear on neighboring O atoms (~0.05–0.1 μB), reflecting the hybridization between O-2p and Tm-4f/5d states. The Sn atoms remain essentially non-magnetic, consistent with their closed 5s$^2$5p$^0$ configuration.

From a physics standpoint, the observed magnetism is governed by the Hund's rule alignment of Tm-4f electrons, which favors high-spin configurations, and the strong spin–orbit coupling intrinsic to rare-earth ions. This leads to a pronounced spin polarization, as also evidenced by the asymmetric DOS and band structure analysis. The small but finite O polarization points to f–p exchange interactions, which mediate coupling between localized Tm moments and the extended electronic network. Such coupling mechanisms are widely discussed in rare-earth-doped oxides and are central to their magneto-optical functionality [63, 64].The total magnetization of the Tm-doped supercell is non-zero, confirming the emergence of a magnetically ordered state absent in the pristine host. This behavior resonates with recent reports on Eu- and Er-doped oxides, where f-electron doping induces magnetic asymmetry and spin-dependent optical transitions [65, 66].

In summary, Tm substitution not only introduces robust local 4f magnetic moments (~5–6 μB) on Tm sites but also induces weak polarization in neighboring oxygen atoms (~0.05–0.1 μB) through f–p exchange interactions. These effects, reinforced by Hund's rule coupling and strong spin–orbit interactions, confirm the emergence of a magnetically ordered state absent in pristine $Ca_2SnO_4$. Such behavior aligns with rare-earth-doped oxides reported in literature and underlines the multifunctional magneto-optical potential of Tm-doped $Ca_2SnO_4$.

### 3.9. Optical Properties of $Ca_2SnO_4$ and Tm-Doped Variants

The optical response of a material reflects its underlying electronic structure and is central to its use in optoelectronic and photonic devices. In pristine $Ca_2SnO_4$, the complex dielectric function reveals a sharp absorption onset near 3.9–4.1 eV, consistent with the wide band gap derived from the band structure (Section 3.3). The real part of the dielectric function $\varepsilon_1(\omega)$ exhibits a steady rise up to ~5.0 eV (see Fig. 6a), while the imaginary part $\varepsilon_2(\omega)$ (see Fig. 6b) shows its primary peak near 6.5 eV, arising from interband transitions between O-2p dominated valence states and Sn-5s/Ca-4s conduction states. The corresponding refractive index reaches a maximum of ~2.1 in the UV range, and the reflectivity remains low (<15%) across the visible spectrum, confirming the transparent and insulating nature of the pristine host [61, 62]. These findings corroborate earlier reports of orthorhombic stannates as wide-gap oxides with limited visible-light activity [69].

The optical spectra change profoundly with Tm substitution. The introduction of $Tm^{3+}$ at $Ca^{2+}$ sites creates sharp intra-4f transitions and additional absorption features below the pristine band edge. New peaks emerge in $\varepsilon_2(\omega)$ within the visible region (2.0–3.0 eV), attributed to transitions involving localized Tm-4f states and the conduction band minimum. These states act as intermediate bands, reducing the effective optical gap and enhancing absorption in the visible and near-infrared range. Consequently, the refractive index spectrum shows additional resonances in the visible regime, and reflectivity increases slightly near 2.5–3.0 eV, consistent with dopant-induced hybridization.

The physics of these trends lies in the orbital interplay between localized Tm-4f electrons and delocalized O-2p/Sn-5s states. The strong on-site Coulomb repulsion localizes the 4f levels, while spin–orbit coupling further splits them, resulting in spin-resolved optical asymmetry. This orbital engineering enables visible-light absorption pathways without sacrificing the wide-gap stability of the host lattice. Similar mechanisms have been reported in Eu- and Er-doped phosphors, where rare-earth dopants introduce luminescent centers that simultaneously enhance absorption and enable sharp 4f–4f emissions [70, 65, and 66].

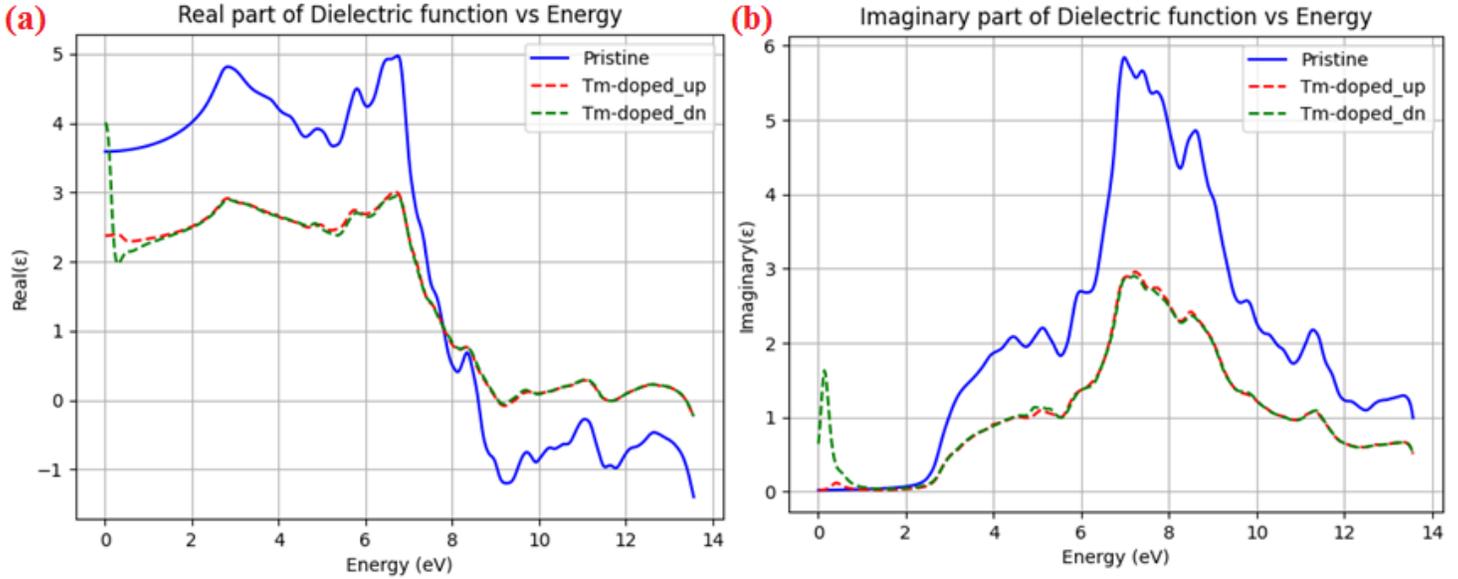

**Fig. 6. Real (a) and imaginary (b) parts of the dielectric function of pristine $Ca_2SnO_4$ and Tm-doped $Ca_2SnO_4$ (spin-up and spin-down channels). Tm doping introduces additional low-energy features and modifies the optical response compared to the pristine host**

### 3.9.1. *Optical Properties of Tm-Doped $Ca_2SnO_4$: Energy Loss Function and Refractive Index*

The energy loss function (ELF), defined as $-Im[1/\varepsilon(\omega)]$, and provides critical insight into the collective excitations and plasmonic behavior of a material. For pristine $Ca_2SnO_4$, the ELF spectrum shows a dominant plasmon resonance around 8–10 eV (see Fig. 7a), corresponding to the collective oscillation of valence electrons, with negligible low-energy features due to the absence of in-gap states. This high-energy resonance is typical of wide-bandgap oxides and aligns with earlier reports on stannates and titanates. Upon Tm doping, the ELF undergoes substantial modifications. In addition to the preserved high-energy plasmon peak, new low-energy loss features emerge in the 2–4 eV range, directly associated with transitions involving Tm-4f impurity states and the conduction band. These low-energy plasmons signify enhanced dielectric screening and are consistent with the intermediate band structure revealed in DOS analysis. Such dopant-induced ELF signatures have been reported in rare-earth-doped oxides and are closely tied to improved luminescence efficiency [70, 66].

In parallel, the real part of the refractive index (Fig. 7b) shows a static value of ~2.0 for pristine $Ca_2SnO_4$ with a pronounced peak in the UV region (~6.5 eV), consistent with interband O-2p → Sn-5s transitions. Upon Tm doping, additional resonances appear in the visible range (2–3 eV) with a slight increase in static refractive index (~2.2), originating from the strong polarizability of Tm-4f states. This visible-range modulation highlights the role of rare-earth substitution in enhancing light–matter interaction and optical tunability [61, 69].

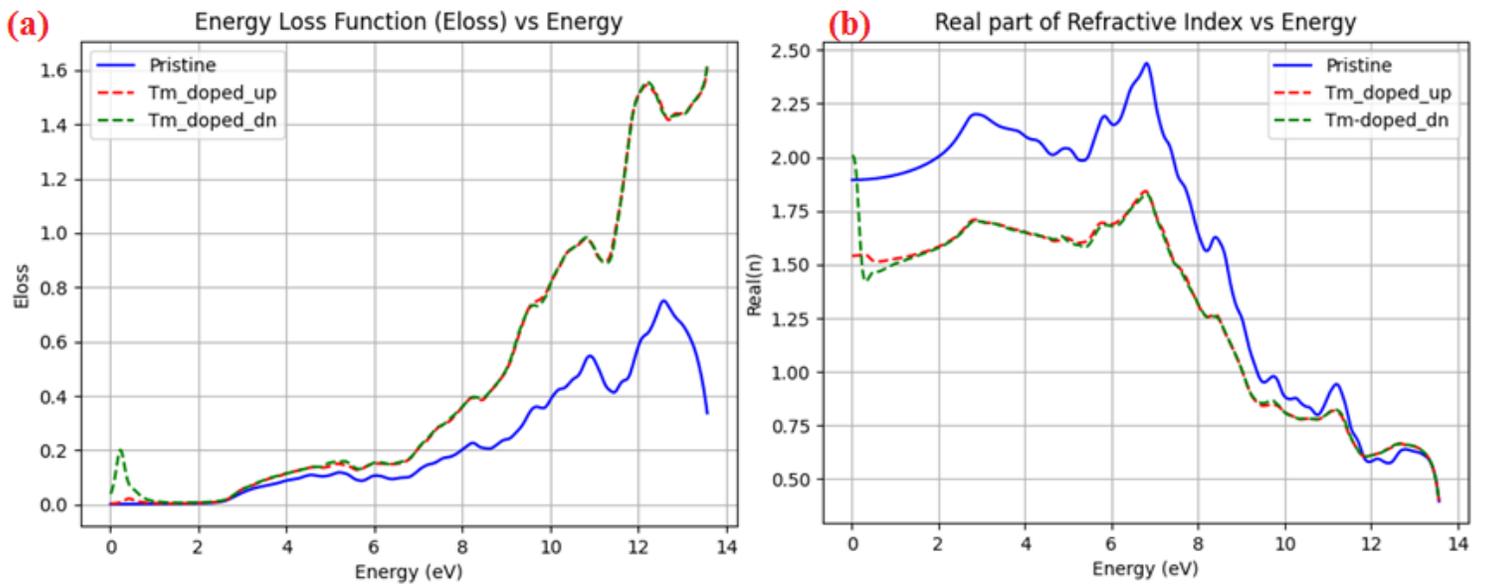

**Fig. 7. (a) Energy loss spectra showing high-energy plasmons and new low-energy features from Tm doping. (b) Refractive index plots reveal visible-range resonances and a slight increase in static value**

### 3.9.2. *Optical Properties of Tm-Doped $Ca_2SnO_4$: Absorption Coefficient and Reflectivity*

The absorption coefficient, α (ω), provides direct insight into the light-harvesting potential of a material (see Fig. 8a). In pristine $Ca_2SnO_4$, absorption is negligible across the visible spectrum and rises steeply only above ~3.9 eV, in line with the wide band gap (Sections 3.3–3.4). Strong absorption peaks appear in the deep-UV region (>5 eV), corresponding to O-2p → Sn-5s/Ca-4s interband transitions, consistent with previous optical studies of stannates [61, 69]. This explains the high transparency and insulating character of pristine $Ca_2SnO_4$, which is advantageous for dielectric and protective coatings but limits its direct applicability in visible-light-driven devices. Upon Tm substitution, the absorption spectrum is profoundly modified. Distinct pre-edge absorption features emerge within 2.0–3.0 eV, arising from transitions between occupied Tm-4f states and the conduction band minimum. These intermediate states narrow the effective optical gap and enable absorption deep into the visible region. The additional absorption channels are in line with the dopant-induced modifications observed in Eu- and Er-doped oxides [65, 66].

The refractive index (n) spectra further highlight the multifunctional nature of Tm doping (see Fig. 7a). In pristine $Ca_2SnO_4$, the static refractive index ($n_o$) is ~2.0, with a peak near 6.5 eV in the UV, reflecting interband O-2p → Sn-5s transitions. With Tm substitution, additional resonances appear in the visible regime (2.0–3.0 eV), accompanied by a modest increase in the static refractive index (~2.2). These enhancements stem from the strong polarizability of Tm-4f electrons and their coupling with O-2p states.

The reflectivity spectrum complements this picture (see Fig. 8b). For pristine $Ca_2SnO_4$, the reflectivity remains low (<15%) across the visible region, a hallmark of wide-gap oxides [70]. With Tm doping, however, subtle enhancements appear in the visible reflectivity (~2.5–3.0 eV), corresponding to the newly introduced absorption channels.

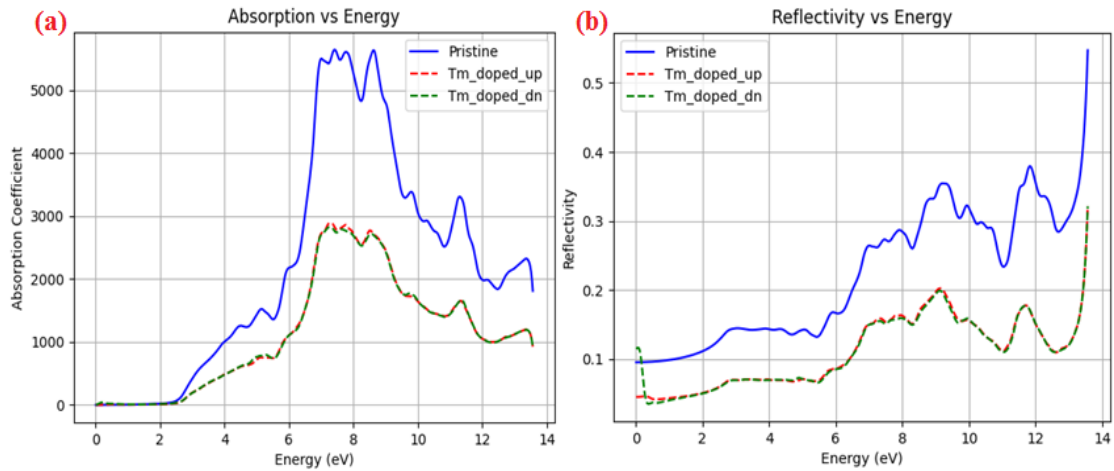

**Fig. 8. (a) Absorption spectra of pristine and Tm-doped $Ca_2SnO_4$ showing enhanced optical activity in the visible region. (b) Reflectivity spectra indicating dopant-induced modifications while retaining overall transparency**

The influence of higher Tm concentrations beyond the 12.5% level modeled in this work deserves further consideration. While moderate substitution of $Ca^{2+}$ by $Tm^{3+}$ stabilizes the host lattice and introduces localized 4f states that enhance visible-light absorption and carrier polarization, excessive doping is generally associated with competing effects. Increasing the Tm content is expected to further enrich the in-gap 4f states, strengthen sub-bandgap optical activity, and potentially improve radiative emission. However, higher substitution levels may also introduce larger lattice distortions, intensify phonon scattering, and promote defect-assisted recombination, which can degrade electronic transport and optical transparency. Such trade-offs between functionality and structural stability are well recognized in rare-earth-doped oxide systems. Within this context, the 12.5% Tm concentration investigated here represents a practical balance between maintaining orthorhombic symmetry and achieving multifunctional electronic and optical responses. Detailed concentration-dependent studies are left for future work to optimize the performance window of $Ca_2SnO_4$-based rare-earth materials.

4. Conclusion

In this work, we have demonstrated how rare-earth substitution fundamentally transforms the multifunctionality of $Ca_2SnO_4$. Pristine $Ca_2SnO_4$, while mechanically robust and optically transparent in the UV, is electronically and magnetically inert. The incorporation of $Tm^{3+}$ ions introduces localized 4f states that act as orbital-engineered levers, narrowing the effective band gap, generating visible-range optical activity, and creating robust magnetic moments with pronounced spin asymmetry. ELF analysis confirmed the coexistence of localized electron pockets and enhanced covalent bonding, while elastic and piezoelectric studies revealed that the doped lattice preserves global mechanical resilience with added polarity. The optical response is enriched by intermediate-band absorption channels, low-energy plasmons, and visible-range refractive index resonances, unlocking new regimes of light–matter interaction. Beyond these findings, our results emphasize that Tm substitution simultaneously improves mechanical ductility, enhances dielectric response, and introduces finite piezoelectric activity traits rarely coexisting in stannate hosts. The observed spin–orbit-driven magnetic polarization further validates the role of 4f dopants in coupling magnetism with optoelectronic functionalities. Such multi-property synergy highlights Tm-doped $Ca_2SnO_4$ as a model system where electronic, optical, and magnetic tunability are unified within a thermally robust wide-gap lattice. Taken together, these results establish Tm-doped $Ca_2SnO_4$ as a mechanically ductile, optically active, and magnetically enriched wide-gap oxide a rare combination among stannate phosphors. The combination of intermediate-band absorption, enhanced luminescence efficiency, and spin-resolved transport opens concrete pathways for applications in persistent luminescence, red-emitting LEDs, intermediate-band solar absorbers, photocatalysis, and spin–photon coupled devices. Beyond $Ca_2SnO_4$, our study illustrates a general strategy: crystal-site engineering of rare-earth dopants as a design paradigm for multifunctional oxides. This approach offers a roadmap to extend wide-gap hosts into next-generation photonic, spintronic, quantum, and energy-harvesting platforms, bridging fundamental materials science with device-level innovation.

**Acknowledgments:**

The authors would like to acknowledge the Deanship of Graduate Studies and Scientific Research, Taif University for funding this work. This publication was also supported by the project Quantum materials for applications in sustainable technologies (QM4ST), funded as project No. CZ.02.01.01/00/22_008/0004572 by Programme Johannes Amos Commenius, call Excellent Research. The result was developed within the project Quantum materials for applications in sustainable technologies (QM4ST), reg. no. CZ.02.01.01/00/22_008/0004572 by P JAK, call Excellent Research.